\documentclass[10pt,draft]{dis03}
\usepackage{epsf,amsmath}

\newcommand{\ie}{{\it i.e.}}
\newcommand{\eg}{{\it e.g.}}

\newcommand{\wrt}{{\it {wrt. }}}

\newcommand{\bs}{\boldsymbol}
\newcommand{\qu}{{\rm q}}
\newcommand{\qb}{${\rm\bar q}$}
\newcommand{\qbm}{{\rm\bar q}}
\newcommand{\qq}{\qu\qb\ }

\newcommand{\lqcd}{\Lambda_{QCD}}
\newcommand{\ieps}{i\varepsilon}

\newcommand{\eq}[1]{(\ref{#1})}

\newcommand{\beq}{\begin{equation}}
\newcommand{\eeq}{\end{equation}}
\newcommand{\beqa}{\begin{eqnarray}}
\newcommand{\eeqa}{\end{eqnarray}}
\newcommand{\nn}{\nonumber}

\newcommand\npb[3]{{Nucl.\ Phys.\ }{\bf B#1}{ (#3)} { #2}}
\newcommand\plb[3]{{Phys.\ Lett.\ }{\bf B#1}{ (#3)} { #2}}

\newcommand\prd[3]{{Phys.\ Rev.\ }{\bf D#1}{ (#3)} { #2}}
\newcommand\prep[3]{{Phys.\ Rept.\ }{\bf #1}{ (#3)} { #2}}

\begin{document}

\title{RESCATTERING EFFECTS ON DIS PARTON DISTRIBUTIONS\thanks{Talk presented at DIS 2003, St. Petersburg, April 2003. Research supported in part by the European Commission under contract HPRN-CT-2000-00130, and by grant no. 102046 of the Academy of Finland.}}

\author{Paul Hoyer\\
        Department of Physical Sciences and Helsinki Institute of Physics\\
        POB 64, FIN-00014 University of Helsinki, Finland \\
\\}

\maketitle

\vspace{-3.3cm}
\hbox to\hsize{\normalsize\hfil\rm HIP-2003-42/TH}
\vspace{2.5cm}
\begin{abstract}
\noindent Rescattering of the struck quark in Deep Inelastic Scattering implies that measured parton distributions are not directly related to the Fock state probabilities of the target wave function. The production amplitudes acquire dynamical phases, which gives rise to shadowing and diffraction in DIS.
\end{abstract}

\section{Rescattering in DIS}

Deep Inelastic Scattering, $e+ N \to e + X$, is one of our most precise tools for investigating the substructure of hadrons and nuclei. According to the QCD factorization theorem \cite{css} the DIS cross section is given by gauge invariant {\em parton distributions}, \eg, for the quark,
\beqa
f_{\qu/N}(x_B,Q^2)&=& \frac{1}{8\pi} \int dx^- \exp(-ix_B p^+ x^-/2)
\label{melm}\\
&\times&\langle N(p)| \qbm(x^-) \gamma^+\, [x^-,0] \qu(0)|N(p)\rangle \nonumber
\eeqa
where the path ordered exponential (POE) is
\beq \label{poe}
[x^-,0] \equiv {\rm P}\exp\left[\frac{ig}{2}\int_0^{x^-}dw^- A^+(w^-) \right]
\eeq
All fields are evaluated at a relative Light-Cone (LC) time $x^+ =t+z \sim 1/\nu$ and
transverse separation $x_\perp \sim 1/Q$, which vanish in the Bjorken limit.

The POE is often ignored since $[x^-,0]=1$ in LC gauge ($A^+=0$). The parton distribution then turns into an overlap of the target Fock state wave functions $\psi_n$ defined at equal $x^+$, \ie, into the {\em probability density}
\beq 
\mathcal{P}_{\qu/N}(x_B,Q^2)= \sum_n
\int^{k_{i\perp}^2<Q^2}\left[ \prod_i\, dx_i\, d^2k_{\perp
i}\right] |\psi_n(x_i,k_{\perp i})|^2 \sum_{j=q} \delta(x_B-x_j)
\label{dens} 
\eeq

In this talk I shall discuss why the parton distribution $f_{\qu/N}(x_B,Q^2)$ actually cannot be reduced to the probability density $\mathcal{P}_{\qu/N}(x_B,Q^2)$ \cite{bhmps}. The physical, intuituive reason is simple: the struck quark rescatters on the color field of the target (Fig. 1a). This rescattering is described by the path ordered exponential \eq{poe} and affects the parton distributions -- when the rescattering occurs within the coherence (or `Ioffe') length of the virtual photon,
\beq \label{ioffe}
L_I = \frac{1}{Q}\cdot\frac{\nu}{Q}= \frac{1}{2m_Nx_B}
\eeq
Rescattering in the final state has important physical effects since it gives rise to kinematically {\em on-shell intermediate states} and hence to complex phases and interference between the rescattering amplitudes. Quantum mechanical interference is crucial for understanding the diffraction, shadowing and polarization effects which are prominently observed in DIS. The nucleon Fock amplitudes $\psi_n$ in \eq{dens} have no dynamical phases since their intermediate states are off-shell.

\begin{figure}[bt]
\centerline{\epsfxsize=10cm\epsfbox{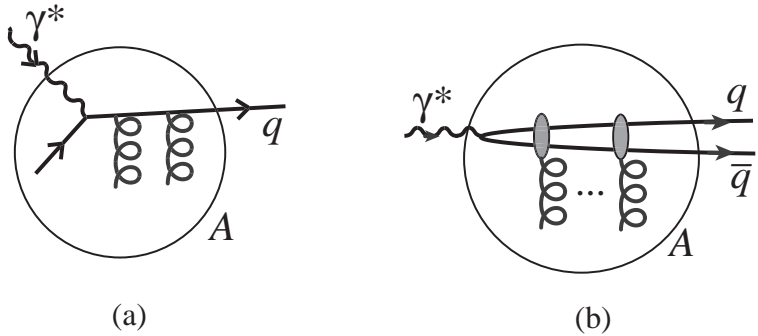}}   
\caption{LC time $(x^+=t+z)$ ordered dynamics of deep inelastic scattering. (a) The DIS frame (6), where the virtual photon momentum $q^z \simeq -\nu$. The photon hits a target quark, which Coulomb rescatters before exiting the target $A$. The increase in $t$ is compensated by a decrease of $z$ such that the photon probes the target at an instant of $x^+$. (b) The Dipole frame (7), where $q^z \simeq +\nu$. The photon splits into a \qq pair at LC time $x^+ \sim 1/2m_N x_B$ before the target. The DIS cross section is given by the scattering of the \qq color dipole in the target.
 \label{fig1}}
\end{figure}

These physical effects of the POE \eq{poe} obviously must remain in any gauge. The reason that the parton distribution \eq{melm} does not reduce to the probability density \eq{dens} stems from the fact that the gluon propagator in LC gauge,
\beq 
d_{LC}^{\mu\nu}(k) =
\frac{i}{k^2+\ieps}\left[-g^{\mu\nu}+\frac{n^\mu k^\nu+ k^\mu
n^\nu}{n\cdot k}\right]  \label{lcprop} 
\eeq
has a pole at $n\cdot k = k^+ =0$. Coulomb scattering of the struck quark on its way out of the target involves momentum transfers $k^+ \sim 1/\nu$ which makes LC gauge singular in the Bjorken limit. This, in turn, implies that interactions between the spectator partons in the target cannot be ignored. The `frozen target' approximation, according to which interactions among the spectators can be ignored during the coherence time of the hard $\gamma^*$ interaction thus fails in LC gauge.

Briefly, the reason for the failure of the frozen target approximation is as follows. The $k^+=0$ poles are gauge artifacts, hence the sum of their residues must add up to zero in the complete, gauge invariant scattering amplitude. This cancellation involves diagrams with gluon exchanges to the struck quark (`rescattering diagrams') as well as exchanges between spectators. The reduction of the path ordered exponential \eq{poe} to unity in $A^+=0$ gauge occurs through a cancellation\footnote{Depending on the prescription used at $k^+=0$ this cancellation can occur separately for each Feynman diagram, or only in their combined contribution to the DIS cross section, as explained in Refs. \cite{bhmps,Belitsky:2002sm}.} between the $-g^{\mu\nu}$ (Feynman) and $(n^\mu k^\nu+k^\mu n^\nu)/k^+$ (LC gauge artifact) parts of the propagator \eq{lcprop}, for rescattering of the struck quark.

Once the $k^+=0$ poles of the LC gauge propagator have been used to cancel the Feynman gauge contribution in the rescattering diagrams the remaining spurious poles in the spectator system give a non-vanishing contribution to the leading twist DIS cross section. In fact, their sum must equal the contribution from the path ordered exponential in Feynman gauge. Thus gauge independence is achieved, and the simplification of the exponential ({\it alias} rescattering of the struck quark) in LC gauge is accompanied by a complication in the spectator system.

\section{Two views of DIS from the target rest frame}

According to \eq{ioffe} the coherence length is long at low $x_B$. This is also the region where shadowing and diffraction phenomena set in. In the following I shall therefore have low $x_B$ in mind -- but the general conclusions are valid at any $x_B$. 

There are two essentially different frames for viewing DIS dynamics, often referred to as the ``Infinite Momentum Frame'' and ``Target Rest Frame'' (see \cite{Collins:2001hp} for a discussion). Actually both views may be had in the target rest frame (LC physics is invariant under longitudinal boosts). The crucial difference is whether the photon moves in the negative or positive $z$-direction. The description in terms of target LC wave functions defined at equal LC time $x^+$ is quite different in the two cases.

\subsection{DIS frame}

I shall refer to the target rest frame where the photon moves along the negative $z$-axis as the {\em DIS frame}: $q = (\nu,\bs{0}_\perp,-\sqrt{\nu^2+Q^2})$. We have then 
\beq \label{qdis}
q^- \simeq 2\nu,\hspace{1cm} q^+ = -Q^2/q^- \simeq -m_N x_B <0\ \  \mbox{ and }\ \  q_\perp=0
\eeq
 Since the photon moves essentially with the velocity of light, it traverses the target in vanishing LC time, $x^+ = t+z \simeq 1/\nu$, even though both $t$ and $z$ are of $\mathcal{O}$(fm). This is why the DIS frame gives a snapshot of the proton wave function at a given $x^+$.

In $x^+$ ordered perturbation theory all 4-momenta are treated as on-shell, and the ``plus'' components of forward-moving particles are positive\footnote{The virtual photon is here regarded as an external particle with negative squared mass $q^2=-Q^2$ and thus, exceptionally, $q^+ < 0$ in \eq{qdis}.}. The conservation of ``plus'' momentum then forbids the transition $\gamma^*(q) \to \qu\qbm$ for $q^+<0$. Thus we have only the LC time ordering $\gamma^*\qu \to \qu$ of Fig. 1a: the photon scatters on a target quark, revealing target structure. 

\subsection{Dipole frame}

In  the {\em Dipole frame} the target is also at rest, but the photon moves along the positive $z$-axis: $q = (\nu,\bs{0}_\perp,+\sqrt{\nu^2+Q^2})$. This frame is related to the DIS frame by a $180^{\circ}$ rotation around a transverse axis. Such a rotation is a dynamical transformation for a theory quantized at $x^+=0$. Hence there is no simple relation between the DIS and dipole frames in the case of non-perturbative target structure (the two frames can obviously be connected for a perturbative target model). The kinematics is inverted \wrt that of \eq{qdis},
\beqa \label{dipcoh}
q^+ \simeq 2\nu \hspace{2cm}&& x^- \sim \frac{1}{2\nu} \to 0 \nn \\
q^- \simeq -\frac{Q^2}{2\nu} \hspace{2cm} && x^+ \sim \frac{1}{m_N x_B} 
\eeqa
Hence in the dipole frame the scattering occurs over a finite LC time $x^+ \sim L_I$. Since $q^+$ is positive there can be a $\gamma^* \to \qu\qbm$ transition before the interaction in the target (Fig. 1b). The $\qu\qbm$ forms a color dipole whose target cross section is determined by its transverse size. In the aligned jet kinematics (which corresponds to the parton model at the lowest order of QCD) the quark takes nearly all the longitudinal momentum, $p_\qu^+ \simeq q^+ \simeq 2\nu$, while $p_\qbm^+ \sim \lqcd^2/m_Nx_B$. The transverse separation of the quark pair is then $r_\perp(\qu\qbm) \sim 1/\lqcd \sim 1$ fm \cite{Piller:1999wx}. Such a transversally large \qu\qb\ dipole has a non-perturbative cross section, which is the parameter in the dipole frame that corresponds to the parton distribution of the DIS frame.

The \qu\qb\ pair multiply scatters in the target as shown in Fig.~1b. At low $x_B$ also the antiquark momentum is large and one expects pomeron exchange, \ie, diffractive DIS, as well as shadowing of the target. The dipole frame is natural for modelling these features of the data, where phases and interference effects play a key role \cite{Piller:1999wx}.

\section*{Acknowledgments} This talk is based on work done with Stan Brodsky, Nils Marchal, St\'ephane Peign\'e and Francesco Sannino \cite{bhmps}.

\end{document}